\documentclass[final]{aipproc}

\layoutstyle{6x9}

\usepackage{amsmath,amsfonts,amssymb}

\allowdisplaybreaks

\begin{document}

\title{Measurement of $\theta_{13}$ by reactor experiments}

\author{Osamu Yasuda\thanks{Based on the work \cite{Minakata:2002jv}
in collaboration with
H.~Minakata, H.~Sugiyama, K.~Inoue and F.~Suekane.
Talk presented at the 5th International Workshop on
Neutrino Factories \& Superbeams (NuFact'03), Columbia University,
New York, USA, June 5-11, 2003.}}{
address={Department of Physics, Tokyo Metropolitan University
1-1 Minami-Osawa,
Hachioji, Tokyo 192-0397, Japan}
}

\begin{abstract}
I describe how
reactor measurements of $\sin^22\theta_{13}$ 
can be improved
by a near-far detector complex.
I show that in the Kashiwazaki plan
it is potentially possible to measure
$\sin^2{2\theta_{13}}$ down to 0.02.
\end{abstract}

\maketitle

{\it Introduction.}
After the successful experiments on atmospheric and
solar neutrinos and KamLAND, the next step in neutrino
oscillation physics is to determine $\theta_{13}$.
It has been known
that the oscillation parameters
$\theta_{jk}$, $\Delta m^2_{jk}$, $\delta$ cannot be determined uniquely
even if the appearance probabilities
$P(\nu_\mu\rightarrow\nu_e)$ and
$P(\bar{\nu}_\mu\rightarrow\bar{\nu}_e)$
are measured precisely from a long
baseline accelerator experiment
due to so-called parameter degeneracies,
and several ideas have been proposed to solve the problem.
Among others, combination of a reactor measurement and a long
baseline experiment offers a promising 
possibility \cite{Fogli:1996pv,Barenboim:2002nv,Minakata:2002jv,Huber:2003pm}.
In this talk I briefly explain
how
measurements of $\sin^22\theta_{13}$ 
in reactor experiments can be improved
by a near-far detector complex.
I also show that in the Kashiwazaki plan \cite{Suekane:2003nh}
the sensitivity to $\sin^22\theta_{13}$ is approximately 0.02.

{\it Reactor measurements of $\theta_{13}$.}
In the three flavor framework
the disappearance probability of the reactor
neutrinos is given by
\begin{eqnarray}
P(\bar{\nu}_{e} \rightarrow \bar{\nu}_{e}) =1-
\sin^22\theta_{13}
\sin^2\left(\frac{\Delta m^2_{13}L}{4E}\right),
\nonumber
\end{eqnarray}
if the contribution from $\Delta m^2_{21}$ is negligible.
So the analysis is reduced to that of the conventional two flavor
framework in vacuum.

Let me start with the derivation of $\chi^2$ used
in \cite{Minakata:2002jv} (See also \cite{multir}).
For simplicity I assume one
reactor and two detectors at near and far distances.
$\chi^2$ is defined as
\begin{eqnarray}
\chi^2&\equiv&\min_{\alpha,\alpha^n,\alpha^f}
\left\{\left[\frac{M^n-T^n(1+\alpha+\alpha^n)}{T^n\sigma_{stat}^n}
\right]^2
+\left[\frac{M^f-T^f(1+\alpha+\alpha^f)}{T^f\sigma_{stat}^f}
\right]^2\right.\nonumber\\
&{\ }&\qquad\qquad\left.+\left(\frac{\alpha}{\sigma_c}\right)^2
+\left(\frac{\alpha^n}{\sigma_u}\right)^2
+\left(\frac{\alpha^f}{\sigma_u}\right)^2
\right\},
\label{chi1}
\end{eqnarray}
where the superscripts $n$ and $f$ stand for the quantities at
the near and far detectors, $M$ and $T$ stand for the measured
and theoretical total numbers of events,
$\sigma_{stat}^n=(T^n)^{-1/2}$ and $\sigma_{stat}^f=(T^f)^{-1/2}$
stand for the statistical errors,
and $\alpha$, $\alpha^n$, $\alpha^f$ are the variables to introduce
the correlated systematic error $\sigma_c$ and the uncorrelated
systematic error $\sigma_u$
(I am assuming that the uncorrelated errors for the two detectors are
the same).  It is understood that the right hand side of Eq.
(\ref{chi1}) is minimized with respect to the three variables.
After eliminating them I get
\begin{eqnarray}
\chi^2=
\left(
\begin{array}{cc}
y^n,&y^f
\end{array}\right)
\left(
\begin{array}{ll}
\sigma_c^2+\sigma_u^2+(\sigma_{stat}^n)^2&\sigma_c^2\\
\sigma_c^2&\sigma_c^2+\sigma_u^2+(\sigma_{stat}^f)^2
\end{array}\right)^{-1}
\left(\begin{array}{c}
y^n\\
y^f
\end{array}\right),
\label{chi2}
\end{eqnarray}
where I have defined $y^n\equiv (M^n-T^n)/T^n$,
$y^f\equiv (M^f-T^f)/T^f$.
When the statistical errors are negligible, the error
matrix in Eq. (\ref{chi2}) is easily diagonalized and
is expressed as
\begin{eqnarray}
\chi^2=\frac{1}{2\sigma_u^2}(y^f-y^n)^2
+\frac{1}{2\sigma_u^2+4\sigma_c^2}(y^f+y^n)^2.
\label{chi3}
\end{eqnarray}
The uncorrelated error $2\sigma_u^2$ in Eq. (\ref{chi3})
is the sum of the contributions from the two detectors
and $\sigma_{rel}\equiv \sqrt{2}\sigma_u$ is
referred to as the relative normalization error in
\cite{Declais:1994su}.
Assuming that one can extrapolate the reference values for
the systematic errors of the Bugey experiment \cite{Declais:1994su}
to the CHOOZ detectors \cite{CHOOZ}, the systematic errors
are estimated to be $\sigma_u$=0.8\%/$\sqrt{2}$=0.6\% and
$\sigma_c$=2.6\%.
Eq. (\ref{chi3}) indicates that the contribution from the
sum $y^f+y^n$ is much smaller than that from the
difference $y^f-y^n$.  This was the reason why the $(y^f+y^n)^2$ term in
$\chi^2$ was ignored in \cite{Minakata:2002jv}.
It should be emphasized that Eq. (\ref{chi3}) shows the advantage of
a near-far detector complex, since the correlated systematic
error $\sigma_c$ is canceled in the denominator of
the $(y^f-y^n)^2$ term \footnote{The Krasnoyarsk proposal
\cite{krasnoyarsk} also takes advantage
of a near-far detector complex.}.

\begin{figure}
\includegraphics[scale=0.57]{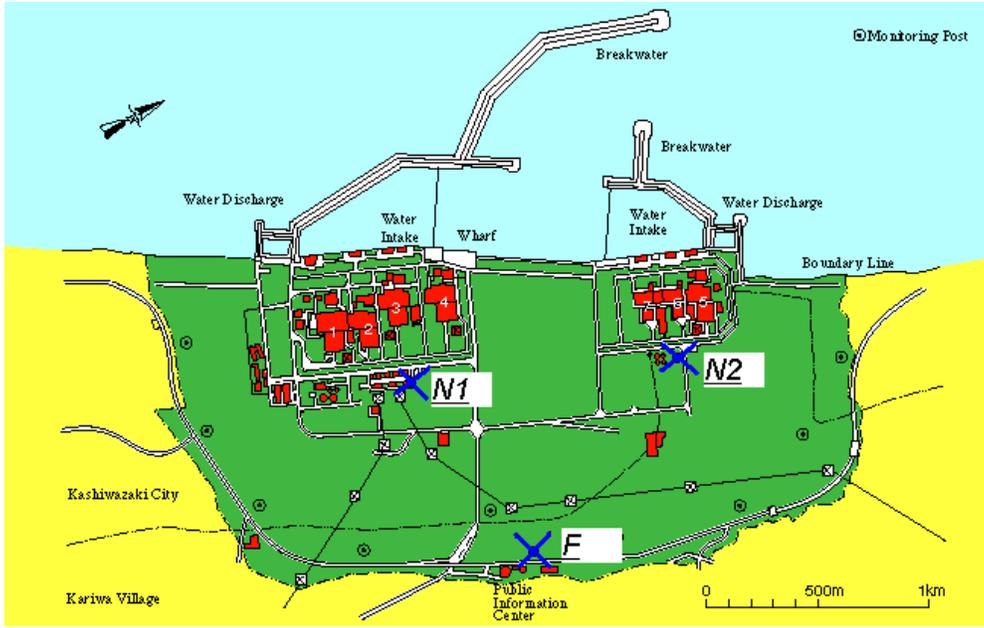}
\caption{\label{fig:kk} The location of the three
detectors in the Kashiwazaki plan.}
\end{figure}

\begin{figure}
\includegraphics[scale=0.7]{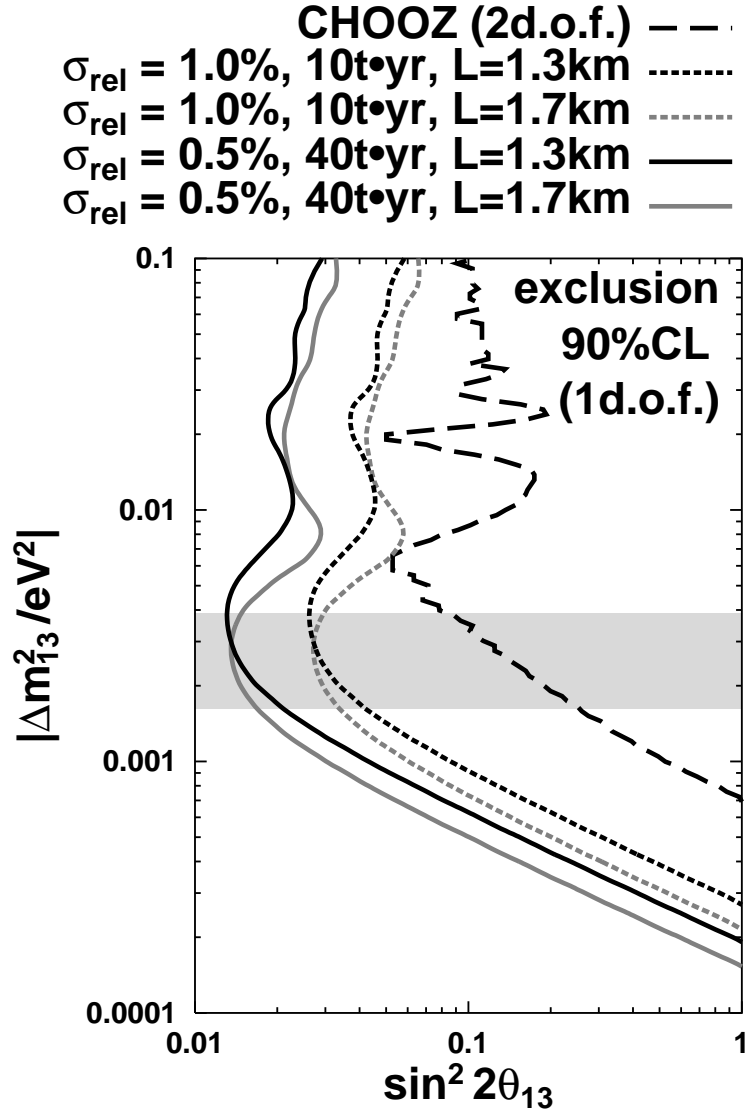}
\caption{\label{fig:exclude} The 90\%~CL\ exclusion limits on
$\sin^2{2\theta_{13}}$ in the Kashiwazaki plan.  The light
shadowed band is the allowed region at 90\%CL for $|\Delta m^2_{13}|$
from the atmospheric neutrino data.}
\end{figure}

From the expression (\ref{chi3}) of $\chi^2$ for the rate,
let me define the following $\chi^2$ for the spectrum
analysis:
\begin{eqnarray}
\chi^2=\sum_j\frac{1}{\sigma_j^2}\left(
\frac{M_j^f-T_j^f}{T_j^f}-\frac{M_j^n-T_j^n}{T_j^n}
\right)^2,
%\label{chi4}
\nonumber
\end{eqnarray}
where $M_j^{n,f}$ and $T_j^{n,f}$ stand for the
measured and expected number of events at the
near and far detectors for the $j$-th bin, and
$\sigma_j$ is the statistical error plus
the uncorrelated systematic error
for each bin:
$\sigma_j^2=1/T_j^n+1/T_j^f+2(\sigma^{bin}_{u j})^2$.
Here I assume that the uncorrelated systematic error
is the same for all bins: $\sigma^{bin}_{u j}=\sigma^{bin}_u$,
so $\sigma^{bin}_u$ is estimated from the uncorrelated systematic error
$\sigma_{u}$
for the total number of events by
\begin{eqnarray}
(\sigma^{bin}_u)^2 =
\sigma_u^2
\frac{(T^f_{tot})^2}
{\sum_j (T_j^f)^2},
\ \ \
T^f_{tot} \equiv \sum_j T_j^f,
\nonumber
\end{eqnarray}
since the uncertainty squared of the total number of events
is obtained by adding up the
bin-by-bin systematic errors
$(\sigma^{bin}_u)^2 (T^f_j)^2$.
The ratio $\sigma^{bin}_u / \sigma_u$
is approximately 3 in our analysis.
Although the sensitivity to
$\sin^22\theta_{13}$ is optimized at
$L\simeq$1.7km for
$|\Delta m^2_{13}|=2.5\times10^{-3}$eV$^2$ \cite{Minakata:2002jv},
the longest baseline
for the far detector inside the campus of the Kashiwazaki-Kariwa
nuclear power plant turns out to be 1.3km \cite{Suekane:2003nh}
(See Fig.\ref{fig:kk}).
In Fig.~\ref{fig:exclude} the 90\,\%\,CL exclusion limits,
which corresponds to $\chi^2 = 2.7$ for one degree of freedom,
are presented 
for two sets of parameters
(data size, $\sigma_{rel}$)=(\mbox{10\,ton-year}, 1\,\%),
(\mbox{40\,ton-year}, 0.5\,\%) and for two baselines
$L=1.3$km, 1.7km, where $\sigma_{rel}\equiv \sqrt{2}\sigma_u$
was introduced earlier \footnote{For simplicity
I have assumed in the calculation that there is only one reactor
and two detectors, but the sensitivity with this simplification
turns out to be almost the same as that with the exact
calculation with seven reactors and three detectors \cite{multir}.}.
Fig.~\ref{fig:exclude} shows that the sensitivity to
$\sin^2{2\theta_{13}}$ does not decrease very much
for $L=1.3$km and that
it is possible to measure
$\sin^2{2\theta_{13}}$ down to 0.02, provided the quoted values of
the uncorrelated systematic error are realized.

{\it Summary \& Conclusions.}
In this talk I emphasized the
advantage of a near-far detector complex
in reactor measurements of $\sin^22\theta_{13}$.
I showed that it is potentially possible to measure
$\sin^2{2\theta_{13}}$ down to 0.02 in the Kashiwazaki plan.

{\it Acknowledgments.}
I would like to thank Hiroaki Sugiyama for many discussions
on the derivation of $\chi^2$.
This work was supported by Grants-in-Aid for Scientific Research
in Priority Areas No. 12047222 and No. 13640295, Japan Ministry
of Education, Culture, Sports, Science, and Technology.

%\vglue 5cm

\end{document}